\documentclass[11pt,twoside]{article}
\usepackage{macro-fsut-eng}
\usepackage{psfig}
\usepackage{graphicx}
\usepackage{cite}
\usepackage[T1]{fontenc} 

\usepackage{latexsym}
\usepackage{verbatim}

\begin{document}

\vskip 1.0cm
\markboth{K.~Pilkington et al.}{Young Stars and Metals in Simulated 
Disk Galaxies}
\pagestyle{myheadings}

\vspace*{0.5cm}
\title{The Distribution of Young Stars and Metals in Simulated 
Cosmological Disk Galaxies}

\author{K.~Pilkington$^{1,2,3}$, B.K.~Gibson$^{1,2,3}$ and D.H.~Jones$^2$}
\affil{$^1$Jeremiah Horrocks Institute, UCLan, Preston, PR1~2HE, UK\\
$^2$Monash Centre for Astrophysics, Clayton, 3800, Australia\\
$^3$Dept of Phys \& Astro, Saint Mary's Univ, Halifax, B3H~3C3, Canada}

\begin{abstract}
We examine the distribution of young stars associated with the spiral
arms of a simulated L$^\star$ cosmological disk galaxy.  We find 
age patterns orthogonal to the arms which are not inconsistent with
the predictions of classical density wave theory, a view further
supported by recent observations of face-on Grand Design
spirals such as M51.  The distribution of metals within a simulated
$\sim$0.1~L$^\star$ disk is presented, reinforcing the link between
star formation, the age-metallicity relation, and the metallicity
distribution function.
\end{abstract}

\section{Star Formation}
\label{star}

We make use of fiducial L$^\ast$ simulated disk galaxy (\tt g15784\rm)
from the McMaster Unbiased Galaxy Survey (MUGS) (Stinson et~al. 2010).
Our earlier work with this simulation has focused on the temporal 
evolution of its metallicity gradient (Pilkington et~al. 2012a) and
metallicity distribution function (Calura et~al. 2012). Here, we examine
briefly the distribution of recent star formation within the simulation, 
with an emphasis on the location of the young stars with respect to the
simulation's most prominent spiral arm.

A detailed description of the \tt g15784 \rm can be found in
Stinson et~al. (2010) and Pilkington et~al. (2012a); here, we simply 
provide a brief overview of the star formation prescription.  The MUGS
simulations were run with the
gravitational N-Body $+$ SPH code \textsc{gasoline} (Wadsley et~al. 2004).
Stars particles are formed with a user-specificed
efficiency from gas particles when
the latter are sufficiently cool ($<$15000~K), dense 
($>$1~cm$^{-3}$), and taking part in a convergent flow 
($\nabla$$\cdot$$v_i$$<$0). Energy feedback from supernovae adheres to the
Stinson et~al. (2006) blastwave formalism.

\begin{figure}
\begin{center}
\hspace{0.25cm}
\psfig{figure=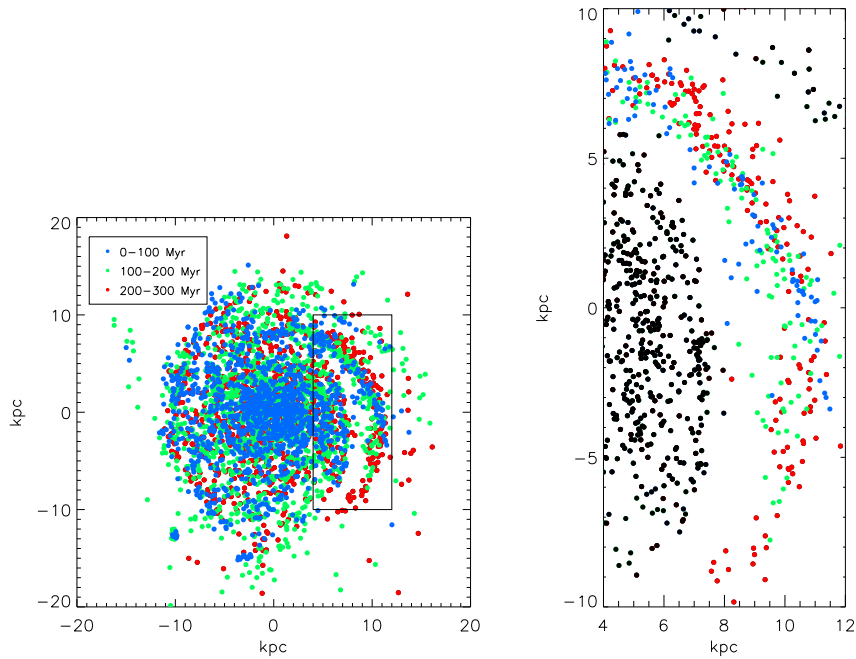,width=14.cm}
\caption{The star particles born in the last 300~Myr (at $z$=0) 
of the \tt g15784 \rm simulation.
The particles are separated into three age bins: `young' (with ages
0$-$100~Myr, shown in blue), `intermediate' (100$-$200~Myr, shown in green), 
and `old' (200$-$300~Myr, shown in red). The black box in the left panel
identifies the most prominent spiral feature, for which an expanded
view is provided in the right panel.  The black points in the latter 
represent the stars born in the last
300~Myr that are not part the spiral arm.}
\label{fig1}
\end{center}
\end{figure}

Figure~\ref{fig1} (left panel)
shows the young stellar population (in this case,
stars born in the last 300~Myr, at redshift $z$=0) of the \tt g15784 \rm
simulation.  The prominence of the centrally-concentrated star formation
has been commented upon already by Stinson et~al. (2010), Pilkington
et~al. (2012a), and Calura et~al. (2012). 
Three 100~Myr age bins are denoted, with the youngest in
blue, the intermediate in green, and the oldest in red.  The right panel
of Figure~\ref{fig1} isolates the most prominent spiral feature
within the simulation (noted by the box inset to the left panel of
Figure~\ref{fig1}). \it We find that the younger (older)
stars tend to lie on the inside/trailing (outside/leading) parts of
the arm. \rm This age `gradient' in the young stellar populations 
orthogonal to the arm is consistent with the basic predictions of 
classical density wave theory, where star formation has been triggered
by gas shocked by the passage of a spiral density wave,
\footnote{At least within co-rotation. We emphasise that
we are {\it not \rm} claiming that this is necessarily what we are witnessing
within the simulation simply that it is consistent with the
predictions of the basic theory.}
see Dobbs \& Pringle (2010).A more thorough
examination of the issues pertaining to spiral arm age gradient/offsets
can be found in Grand et~al. (2012).

The offsets we see in the stellar populations associated with the spiral
arm are most prominent in the `upper' part of the figure, before the 
arm opens up to its full extent. To achieve a fairer comparison with
both the high-resolution models of Dobbs \& Pringle (2010) and the 
observations which show very similar trends (e.g. for M51, 
as in S\'anchez-Gil et~al. 
2011), we would need much finer temporal resolution (where the age
ranges probed are 0$-$10~Myrs, rather than 0$-$300~Myrs); having said that,
the gross trends in orthogonal age gradients/offets do appear to
extend to $\sim$100~Myr old stellar populations (e.g. Calzetti 
et~al. 2005) and so perhaps the result highlight here is not 
obviated by the larger age bins.

\section{Chemical Properties}
\label{chem}

Finally, we wish to show our most recent work on the distribution
of metals in a suite of $\sim$0.1~L$^\star$ disk simulations undertaken
with a wide range of feedback prescriptions (including supernovae and
thermal energy from OB-stars during their pre-supernovae
phase), initial mass functions, and metal diffusion efficiencies.
We focus here on our fiducial simulation, \tt 11mKroupa\rm, which
was first introduced in a different context by Brook et~al. (2012).
This simulation, like \tt g15784\rm, was realized with the 
\textsc{gasoline} code, but with an upgraded version to take into 
account the chemical enrichment histories of broader spectrum of
elements beyond just oxygen and iron.

\begin{figure}
\begin{center}
\hspace{0.25cm}
\psfig{figure=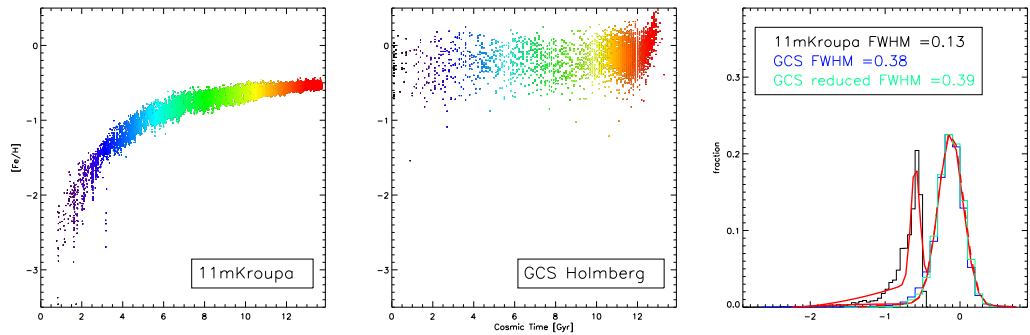,width=14.cm}
\caption{The left-most panel shows
the age-metallicity relation for the stars situated in the 
`solar neighborhood' of the \tt 11mKroupa \rm simulation.
The stars are colored by age where red is the youngest and purple 
is the oldest. The middle panel, for comparison,
shows the observed 
age-metallicity relation for the solar neighborhood of the 
Milky Way (Holmberg et~al. 2009). The right-most panel 
shows the resulting metallicity distribution functions for these
age-metallicity relations, along with simple single Gaussian fits to their
respective `peak' regions (with the FWHM noted in the inset).}
\label{fig2}
\end{center}
\end{figure}

The left-most panel of Figure~\ref{fig2} shows the age-metallicity
relation of the `solar neighborhood' (an 
annulus $\sim$3 disk scalelengths from the center, lying within a kpc
of the mid-plane) associated with \tt 11mKroupa\rm.  The middle
panel shows the the corresponding relationship in the solar neighborhood
of the Milky Way, as derived from the Geneva-Copenhagne Survey
(GCS) by Holmberg et~al. (2009).  The right-most panel shows the 
associated metallicity distribution functions (MDFs) for these 
respective `solar neighborhoods'; the (indistiguishable) 
overlaid curves on the
right-most MDF within the panel correspond to two 'cuts' of
the GCS (effectively, `volume-limited' and `open', in some sense - 
details provided in Pilkington et~al. 2012b), while the left-most MDF is
that constructed from \tt 11mKroupa\rm.

What should be readily apparent from Figure~\ref{fig2} is that the
age-metallicity relation for the solar neighborhood of \tt 11mKroupa\rm
is significantly more correlated than that of the Milky Way's solar
neighborhood and that it is very tight at a given age.  
The former should not be surprising, in that the star formation
and infall histories of the two are not the same.  Regardless, this
tight correlation has an inexorable effect on the resulting MDF, in the
sense that it is more negatively skewed, possesses greater kurtosis, 
while the MDF's `peak' component is quite narrow (due to the minimal
dispersion in [Fe/H] at a given age convolved with the simulation's
star formation history).  A deeper analysis is provided by
Pilkington et~al. (2012b).

\acknowledgments 
Without the help of our collaborators, this work would not have been 
possible; we thank them for their ongoing advice and guidance. We wish 
to thank both Monash and Saint Mary's Universities for their generous 
visitor support.  KP acknowledges the STFC studentship:(ST/F007701/1).


\begin{references}
\reference Brook, C.B., Stinson, G., Gibson, B.K., et~al. 2012, MNRAS, 419, 771
\reference Calura, F., Gibson, B.K., Michel-Dansac, L., et~al. 2012, MNRAS, submitted
\reference Calzetta, D., Kennicutt, R.C., Bianchi, L., et~al. 2005, ApJ, 633, 871
\reference Dobbs, C.L., Pringle, J.E. 2010, MNRAS, 409, 396 {\bf 409 \rm}, 396-404, (2010)
\reference Grand, R.J.J., Kawata, D., Cropper, M. 2012, MNRAS, submitted
\reference Holmberg, J., Nordstr\"om, B., Andersen, J. 2009, A\&A, 501, 941
\reference Pilkington, K., Few, C.G., Gibson, B.K., et~al. 2012, A\&A, in press
\reference S{\'a}nchez-Gil, M.C., Jones, D.H., P\'erez, E., et~al. 2011, MNRAS, 415, 753
\reference Stinson, G.S., Bailin, J., Couchman, H., et~al. 2010, MNRAS, 408, 812
\reference Stinson, G., Seth, A., Katz, N., et~al. 2006, MNRAS, 373, 1074
\reference Wadsley, J.W., Stadel, J., Quinn, T. 2004, New~Astron, 9, 137
\end{references}
\end{document}